# Arterial blood pressure waveform in liver transplant surgery possesses variability of morphology reflecting recipients' acuity and predicting short term outcomes

Shen-Chih Wang[1,2], Chien-Kun Ting[1,2], Cheng-Yen Chen[2,3], Chinsu Liu[2,3], Niang-Cheng Lin[2,3], Che-Chuan Loong[2,3], Hau-Tieng Wu[4,5], Yu-Ting Lin[1,2]

1. Department of Anesthesiology, Taipei Veterans General Hospital, Taipei, Taiwan
2. School of Medicine, National Yang Ming Chiao Tung University, Taipei, Taiwan
3. Division of Transplantation Surgery, Taipei Veterans General Hospital, Taipei, Taiwan
4. Department of Mathematics, Duke University, Durham, NC, US
5. Department of Statistical Science, Duke University, Durham, NC, US

1. hauwu@math.duke.edu
2. linyuting@hotmail.com.tw

# Abstract

Background:
We investigated clinical information underneath the beat-to-beat fluctuation of the arterial blood pressure (ABP) waveform morphology. We proposed the *Dynamical Diffusion Map* algorithm (DDMap) to quantify the *variability of morphology*. The underlying physiology could be the compensatory mechanisms involving complex interactions between various physiological mechanisms to regulate the cardiovascular system. As a liver transplant surgery contains distinct periods, we investigated its clinical behavior in different surgical steps.

Methods:
Our study used DDmap algorithm, based on unsupervised manifold learning, to obtain a quantitative index for the beat-to-beat variability of morphology. We examined the correlation between the variability of ABP morphology and disease acuity as indicated by Model for End-Stage Liver Disease (MELD) scores, the postoperative laboratory data, and 4 early allograft failure (EAF) scores.

Results:
Among the 85 enrolled patients, the variability of morphology obtained during the presurgical phase was best correlated with MELD-Na scores. The neohepatic phase variability of morphology was associated with EAF scores as well as postoperative bilirubin levels, international normalized ratio, aspartate aminotransferase levels, and platelet count. Furthermore, variability of morphology presents more associations with the above clinical conditions than the common BP measures and their BP variability indices.

Conclusions:

The variability of morphology obtained during the presurgical phase is indicative of patient acuity, whereas those during the neohepatic phase are indicative of short-term surgical outcomes.

# Introduction

Blood pressure fluctuates continuously. By characterizing the systolic blood pressure (SBP) or diastolic blood pressure (DBP) variation, the blood pressure variability (BPV) has been positively associated with clinical events[1-3]. While SBP and DBP are merely two measures from the arterial blood pressure (ABP) waveforms, which reflects the integral of the complex pathophysiological interactions in the cardiovascular system[4-7], the beat-to-beat variation of ABP waveform morphology may reveal more dynamical information[8, 9] in comparison of BPV.

To quantify the multivariate ABP waveform dynamics, we used *Dynamic Diffusion Map* (DDMap), an unsupervised learning algorithm, to perform cardiovascular waveform analysis[10, 11]. In liver transplant surgery, we have observed the association of ABP waveform morphologic variation with the recipients' clinical condition (Fig. 1) [12]. It is probably that the complex fluctuation of waveform morphology reflects the versatile interaction of various physiological mechanisms to maintain homeostasis via the regulations of the blood flow, pressure and contractility both in the organ and systemic scales. The more preserved functional capacities for homeostasis a healthier patient has, the more variation in the waveform morphology we can observe, and vice versa. Therefore we hypothesized that the better liver disease condition exhibits more variability of morphology[12-15].

In our study, we first quantified the ABP waveform variation at presurgical and neohepatic phase. Then we examined the association with respect to the clinical scoring systems evaluating liver recipient's preoperative acuity, postoperative laboratory data (POD)[15-19], and early allograft failure (EAF) scores derived from POD. Correlations between these clinical data and common blood pressure (BP) measures, including SBP, DBP, mean BP (MBP), and pulse pressure (PP), were analyzed as well as the BPV indices. We hypothesized that variability of morphology in ABP waveforms implicated more clinical information in comparison to BPV and therefore had stronger correlation to liver recipients' condition.

## Material and Methods

### Patients

This was a single-center prospective observational study conducted at Taipei Veterans General Hospital, Taiwan between 2018 and 2022. A total of 85 living donor liver recipients (LDLR) were recruited after receiving approval from the Institutional Review Board (IRB No.: 2017-12-003CC and 2020-08-005A) and written informed consent from each patient. Note that this study focused on LDLR due to the fact that living donors (i.e., family members) outnumber cadaveric donors in Taiwan[20]. Note also that LDLR is highly beneficial in the context of the current study as it minimizes variation in the cold ischemia time of the graft. The protocol in this study conformed to the ethical guidelines outlined in the 1975 Declaration of Helsinki.

### Liver recipients' acuity

The laboratory data obtained within one week before the surgery were used to calculate the MELD score[21], MELD-Na score[17] and Child-Pugh score[15] for LDLR acuity.

### Physiological data collection

Throughout the liver transplant surgery, continuous physiological waveforms were collected from the patient monitor (GE CARESCAPE™ B850, GE Healthcare, Chicago, IL) with data collection software S5 Collect (GE Healthcare). All physiological data, including electrocardiogram (ECG), ABP, central venous pressure (CVP), and pulse oximetry, were recorded from the beginning of the anesthetic maintenance; however, only ABP was used in subsequent analysis. The recorded waveforms were uniformly sampled at 300 Hz. As a matter of routine, a portocaval shunt was set during the anhepatic phase for the release of portal flow and relief from intestinal congestion. We registered specific time points, including hilum clamping prior to hepatectomy, the end of hepatectomy, portocaval shunt establishment, portal vein reperfusion, completion of hepatic artery reperfusion, and completion of bile duct anastomosis.

Following the operation, the liver recipients were transferred to the intensive care unit (ICU) to undergo postoperative recovery under the discretion of physicians blinded to this study.

### Liver recipients' clinical outcomes

The POD, including levels of aspartate aminotransferase (AST), alanine aminotransferase (ALT), bilirubin, international normalized ratio (INR), and platelet count were checked daily in first 7 days per institutional routine. With these POD, we calculated EAF scores such as MEAF score[22], L-GrAFT$_7$, L-GrAFT$_{10}$, and EASE score [23-25].

### Waveform data preprocessing

We focused data analysis on the presurgical and neohepatic phases. Data based on the recipient's own liver obtained during the presurgical phase, beginning with anesthetic maintenance in preparation for surgery, were used as the baseline condition. We selected the first 800 ABP pulses during the presurgical phase, which accounted for slightly less than 10 minutes of data. The neohepatic phase was identified as a hemodynamically stable 10-minute period following anastomosis of the biliary duct and prior to abdominal wound closure. We assumed that by this point, perfusion and hepatic artery flow would already have been established, such that the new liver graft would already have functioned for a while. Note that the waveforms obtained were carefully checked to ascertain that neither drastic changes nor severe data artifacts exist. The consideration of the constant 10-minute period approach is to ensure applicability to future research. Also, we think 10 minutes is an adequate interval to assess the patient's latest condition from our experience.

In preparing consecutive ABP waveform pulses for DDMap (Fig. 1), we adopted the maximum systolic phase slope as the fiducial point for each pulse cycle[9]. Legitimate ABP pulses were automatically detected and extracted based on pulse width, extreme values, and pulse pressure. Moreover, the duration between two successive pulses should be within a range dynamically determined at the individual level.

### Quantitative variability of morphology

Patterns in the pulse-by-pulse evolution of ABP waveforms are too subtle to be gauged directly by the naked eye or simply measured from raw ABP data (Fig.1 and Fig. S1). Such subtle information could be condensed comprehensively for data visualization and quantification by an advanced unsupervised manifold learning algorithm, DDMap[9, 10]. After the waveform data pre-processing, DDMap converts consecutive pulses into a trajectory of intricate shape. The trajectory provides a concise overview of the complex dynamic evolution of ABP waveforms while the *variability of* *morphology* is quantified as the moving average trend length of the trajectory.

The algorithm[10, 11] achieves the above function via the following steps. First, the algorithm detects the beat-to-beat cardiac cycles in the ABP waveform using a method based on the maximum first derivative. Once all cardiac cycles are identified, the pairwise distances are calculated between the arterial pressure waveform samples of the different cardiac cycles in the selected time-interval. Using a Gaussian function as the kernel function to convert the distance into affinity, we obtain the *affinity matrix* describing the affinities between all pairs of cardiac cycles. The affinity matrix contains dissimilarity information among cardiac cycles. Its eigendecomposition reveals the high-dimensional structure pertaining to all cardiac cycles (wave-shape manifold), from which we can use the time sequence information to establish the waveform morphological dynamics and to calculate its variability. It is unsupervised as the above calculation does not involve any labels, including medical history, lab test data, or details pertaining to the surgical procedure. See Fig. 1 and Fig S1 for an example and see Supplementary for technical details for DDMap and the quantitative variability of morphology.

### The BPV indices

BP measures and variability indices of these BP measures were derived from the corresponding ABP waveforms. According to the relevant previous studies[1, 26], we calculated variability indices, including coefficient of variation (CV), standard deviation (SD), average real variability (ARV), interquartile range (IQR), quartile coefficient of dispersion (QCD), and mean value of SBP, DBP, MBP, and PP to obtain 24 indices for comparison.

### Sensitivity analysis

Several sensitivity analyses were created to assess the quantitative morphological variability index: 1) The choice of parameter; 2) Random perturbations in time sequence labeling were used to examine the ABP waveform complexity of pulse cycle evolution; 3) Perturbations in case labeling of ABP waveforms were used to test our hypothesis that liver disease acuity is evident in pulse cycle evolution; and 4) Perturbations imposed by replacing pulse cycles with mean blood pressure values were used to determine whether the pulse cycle contains information complementary to blood pressure. Finally, the effect of reduced time duration (7 mins and 5 mins) for calculation was investigated.

### Statistical analysis

Demographic and laboratory data and all variability indices related to ABP waveform were expressed as median and interquartile range (IQR). Spearman rank correlation analyses were performed via bootstrap resampling using 100,000 resamples without replacement and reported as mean and the corresponding 95% CI. Linear regression was used to establish the association between the variability of morphology and the liver recipients' acuity before surgery and clinical outcomes. The adjusted $R^2$ in regression analysis and the Spearman rank correlation were also used to compare the variability of morphology with BP variability indices. Statistical significance was indicated by a p value < 0.005 for scientific reproducibility[27].

ABP waveform pre-processing and DDMap implementation were both performed using C# (Visual Studio community 2021, Microsoft, Redmond, WA). All statistical analyses were performed using R, version 4.2.3 (R Foundation for Statistical Computing). Statistical data and the R script used to produce results are provided in the Supplementary Materials. MELD, MELD-Na, and 4 EAF scores were calculated using Excel (Microsoft, Redmond, WA). Note that EAF formulas were sourced from the corresponding authors. Note also that MELD and MELD-Na scores were calculated without rounding off decimals to avoid the round-off errors.

# Results

Our patient data was summed up in table 1. The average age of our patients was 59 [52,62] years old. The main indication for liver transplant was hepatitis B virus-associated hepatocellular carcinoma. The major graft type was right lobe of liver without middle hepatic vein territory reconstruction (51.2%). The average graft to recipient weight ratio was 1.07% [0.88, 1.27]. The 85 enrolled patients presented a wide range of disease acuity (MELD-Na: 20.1 [13.5, 31.6], MELD: 18.2 [10.6, 28.9], Child-Pugh: 10.0 [7.0, 12.0]). Two cases were excluded due to a lack of waveform data in the neohepatic phase resulting from technical issues.

### Association with acuity and the presurgical phase

In the presurgical phase, the baseline variability of morphology (60.1 [48.4, 72.9]) was significantly correlated with MELD (−0.39 [−0.57, −0.18], p=0.0003), MELD-Na (−0.44 [−0.61, −0.24], p=0.00003), and Child-Pugh scores (−0.43 [−0.59, −0.23], p=0.00002). Liver recipients with higher clinical scores had lower variability of morphology. The correlations of the presurgical variability of morphology with all the other epidemiological factors were not significant (Fig.2). It is also worth mentioning that the presurgical variability of morphology was not correlated with 4 EAF scores.

### Association with clinical outcomes and the neohepatic phase

In the neohepatic phase, the variability of morphology (47.0 [42.7,52.1]) was generally lower than those in the presurgical phase (Fig.3); however, correlations with MELD

(−0.40 [−0.60, −0.18], p=0.0004), MELD-Na (−0.43 [−0.62, −0.21], p=0.00005), and Child-Pugh score (−0.35 [−0.55, −0.14], p=0.0012) remained significant (Fig.2). The variability of morphology in the neohepatic phase were also correlated with postoperative laboratory data and surgical outcomes (Fig.2 and 3). In particular, the variability of morphology was lower in those with a higher POD7 TB. The variability of morphology was correlated with the EAF scores, including EASE (−0.32 [−0.52, −0.09], p=0.0033), LGrAFT-7 (−0.32 [−0.52, −0.09], p=0.004), LGrAFT-10 (−0.31 [−0.51, −0.09], p=0.003), and MEAF scores (−0.31 [−0.51, −0.08], p=0.004). Note that the correlations between variability of morphology and other epidemiological factors were not as strong (Fig.2).

### Association with postoperative laboratory data and the neohepatic phase

Using the laboratory data from POD1 to POD7 as a response variable, the neohepatic variability of morphology was the univariate regression factor most strongly associated with TB ($R^2$=0.20, p=0.000024) and DB ($R^2$=0.20, p=0.000018), whereas the correlations with PLT ($R^2$=0.10, p=0.003), INR ($R^2$=0.082, p=0.009), AST ($R^2$=0.034, p=0.093), and ALT ($R^2$=0.016, p=0.25) were insignificance. At POD 5-6, Spearman's correlation (Fig. 4) was most strongly associated with TB and DB (TBday5: -0.383 [-0.180, -0.587], p=0.00016; Tbday6: -0.385[-0.185, -0.559], p=0.00011; Dbday5: -0.437 [-0.238,-0.603], Dbday6: -4.21 [-0.225,-0.588], p<0.00001).

### Predicting variability of morphology based on MELD score and demographic factors

Univariate regression analysis of the presurgical variability of morphology identified MELD-Na as the most significant factor (Coef=−0.65, t=−4.2, p=0.00005), while demographic data including age, gender, and BMI were not significant factors for the presurgical variability of morphology. The same regression analysis for the neohepatic variability of morphology showed that MELD-Na (Coef=0.30, t=−4.3, p=0.00004) and POD7 TB (Coef=−0.54, t=−334, p=0.001) were both significant factors (Adjusted $R^2$=0.30).

### Blood pressure parameter analysis

The variability of morphology performed better than all BP parameters in both correlation and regression analysis with MELD-Na in the presurgical phase and with POD7 TB in the neohepatic phase (Fig.5). Among BP measures and BP variability indices, the IQR of DBP was best correlated with MELD-Na (Coef= -0.31 [-0.50, -0.09], p=0.0031) in the presurgical phase and POD7 TB (Coef= -0.29 [-0.07, -0.36], p=0.006) in the neohepatic phase. The standard deviation of DBP best performed in the regression to MELD-Na (adjusted $R^2$ relative to variability of morphology=0.456), while the average real variability (ARV) of MBP best performed in the regression to POD7 TB (adjusted R2 relative to variability of morphology=0.661).

### Sensitivity analysis

Sensitivity analysis revealed that derivations of the variability of morphology were robust, i.e., insensitive to parameter selection (Supplementary Figs. S2, S3). We also carry out various null models (Fig. S4) to show that the association between the MELD score and presurgical variability of morphology is genuine by randomly shuffling time sequential relationship and labels. Under both null models, the correlations with the MELD score decrease. Reducing time duration for variability of morphology calculation affect performance slightly. The results are detailed in the Supplementary Material.

## Discussion

This was the first study to investigate the variability of morphology in ABP waveforms from patients undergoing liver transplant surgery. While it is difficult to visually summarize human eyes, our results confirmed the hypothesis that the quantitative beat-to-beat variation of waveform morphology implicates clinical information. We showed that it reflects liver recipients' presurgical condition. As for early surgical outcomes, limited by our case number, we could provide only indirect evidence, such as the correlation between variability of morphology and POD bilirubin or EAF scores. To the best of our knowledge, this is also the first study to demonstrate that the ABP waveform dynamics possesses information relevant to liver transplant outcomes. Furthermore, our result suggested that the variation of ABP waveform is more relevant to clinical condition than the common BP measures and the associated BPV indices.

### ABP waveform variability of morphology and presurgical condition

The MELD and MELD-Na scores are used to rank patients awaiting liver transplant in terms of urgency. Note that these scores are also helpful in assessing various liver diseases, which previously relied on the Child-Pugh score.[15] Our results are also consistent with the known fact that conditions prior to transplant surgery are not necessarily indicative of liver transplant surgical outcomes[15, 28]. This situation suggests that during the presurgical period (i.e., after anesthetic induction but before the commencement of surgical procedures), ABP waveforms may provide insight into the baseline condition of the patient. We postulate that patients with a lower (i.e., better) MELD scores possess more homeostatic capability or functional reserve in their cardiovascular system, earning them a higher variability of morphology.

In the presurgical period, the association of variability of morphology with the disease acuity might grant a new perspective for clinicians. The assessment of disease acuity could be achieved with either combining several laboratory measures, or the quantitative beat-to-beat variation of ABP waveform morphology. The signal processing implementation could give rise to future clinical application in 1) timely assessment immediately before the start of surgery, 2) establishment of the individual baseline reference to help subsequent timely evaluation of the condition particularly in the neohepatic phase, as the laboratory examination takes time.

### ABP waveform variability of morphology in the neohepatic phase

Our result shows that ABP waveform morphological variability in the neohepic phase is associated with POD laboratory data, in particular TB, DB and PLT, as well as the 4 EAF scores (L-GrAFT$_7$, L-GrAFT$_{10}$, EASE, and MEAF). This finding substantiates our hypothesis that the morphology variability in the neohepatic phase is indicative of the liver disease status, coinciding with the commencement of functionality in the newly transplanted graft, which exerts positive effects on the regulatory mechanisms underlying the variability of ABP waveform morphology. This result is also in consistent with the development of these clinical scores, including MELD-Na for liver disease acuity assessment and EAF scores for the outcome prediction after the transplant surgery. The fact that MELD and MELD-Na have not proven predictive of graft function recovery has prompted extensive research into graft outcomes. Bilirubin reflects the metabolic function of the liver graft and has been implicated in surgical outcomes[14, 29]. Our POD laboratory data revealed that the neohepatic *variability of morphology* is most associated with TB, followed by the platelet count. The correlation with platelet counts suggests the possible involvement of inflammation or graft regeneration[30, 31]. The lack of a statistically significant correlation between neohepatic variability of morphology and AST or ALT can perhaps be attributed to the minimal ischemia time associated with living donor transplants.

To improve graft outcome prediction, EAF scores were developed in the last decade (e.g., L-GrAFT$_7$, L-GrAFT$_{10}$, EASE, and MEAF) utilizing multiple serial POD laboratory data [23-25]. Our results showed that in the neohepatic phase of surgery, variability of morphology was significantly correlated with these 4 EAF scores [14, 24, 32] and the disease acuity scores (MELD, MELD-Na and Child-Pugh scores). We did not record ABP waveforms after sending patients to ICU. Investigations are needed to elucidate if postoperative ABP waveform variability improves EAF prediction.

It is possible that the conditions of the regulatory physiological mechanisms underlying the variability of morphology carried over from the presurgical to the neohepatic phase, which took merely hours. On the other hand, the surgical influence on the outcome was probably reduced due to the minimal ischemia time and the consistency of the patient management from the same surgical team in this study. Nevertheless, the variability of morphology is in general lower in the neohepatic phase than in the presurgical phase, which probably is caused by the pathophysiological influence of the inevitable stresses of the liver transplant surgery. After harvesting the liver graft, surgeons needed to perfuse the liver graft with ice cold perfusate, mostly histidine-tryptophan-ketoglutarate (HTK) solution. After hepatic and portal vein anastomosis, surgeons reperfused the liver before anastomosing the hepatic artery. The drastic hemodynamic changes elicited by complex vascular procedures, ice cold preservation solution, and the reperfusion injury[13] could negatively impaired the regulating mechanisms and hence the ABP variability of morphology.

Clinical implication

ABP waveforms are commonly used to extract information related to arterial stiffness or imminent shock and to provide hemodynamic profiles during surgery[4-7]. Note however that none of these applications focus on the beat-to-beat evolution of the pulse cycle. In the current study, the variability of morphology was quantified using the proposed method. We were unable to find any similar work in the literature. The one study that most closely resembled our work involved the use of BPV and its correlation with variations in blood pressure[33]. Because the variability of morphology could be obtained at any point during surgery, particularly before the wound closure, which leaves an opportunity for surgical intervention, its potential role in decision-making and if it could further help reduce the EAF rate warrant further investigation.

Methodological considerations

The status of the cardiovascular system is ever-changing. In fact, it has been observed that no two pulses during a surgical procedure could be considered identical in terms of waveform morphology[9]. While the common BP measures are univariate time series, data of BP waveform are *multivariate time series*, which demand sophisticate method to analyze other than simple variance measures in statistics. Our results suggest that the variability of morphology in ABP waveforms closely relative to BP variability (Fig. 5). It is possible that BPV indices capture the partial information of the ABP waveform in beat-to-beat, which is better grasped by the DDmap algorithm based on unsupervised learning for visualizing and quantifying the structure hidden within these waveforms. Since the ABP is routinely monitored during the liver transplant, with the advance of modern monitoring system, the proposed algorithm can be easily integrated into the clinical setup.

### Limitation and Future Works

The current study focused exclusively on living donor liver transplant. Our results are not necessarily representative of cadaveric liver transplant, in light of differences in cold ischemia time. The proposed method should be assessed in the context of cadaveric transplant surgery, preferably in a different institution. Furthermore, our methodology involved the recording of ABP waveforms, which is not usually considered in the recent electronic medical records systems. Thus, most previous studies do not include waveform data, thereby limiting the availability of retrospective data for verification. Third, this is a single institutional study. The reported result should be further verified in a multiple center setup.

Last but not the least, would this intra-operative information ultimately facilitate surgical intervention to improve the outcome? Our current results from the offline data shed light on the potential of variability of morphology. However future studies are needed to elucidate if such potential could facilitate decision making. Regarding the pathophysiology of the variability of morphology, the integration of several regulatory mechanisms and their complex interactions with the clinical environment may prohibit studies under ideally controlled conditions. Future studies may resort to indirect data and advanced statistical techniques.

### Conclusion

The quantitative variability of morphology derived from the pulse cycle evolution in the ABP waveform during the liver transplant surgery reflects the liver disease acuity and is potential to foretell the short-term surgical outcomes during the surgery.

# Acknowledgement

We thank Professor Alfonso Avolio (UCSC, Italy) for kindly providing the details related to the calculation of EASE scores. We also thank Professor Vatche G. Agopian (UCLA, US) for providing details pertaining to the calculations of L-Graft scores.

This study was supported by the National Science and Technology Development Fund (MOST 109-2115-M-075 -001) of the National Science and Technology Council, Taipei, Taiwan.

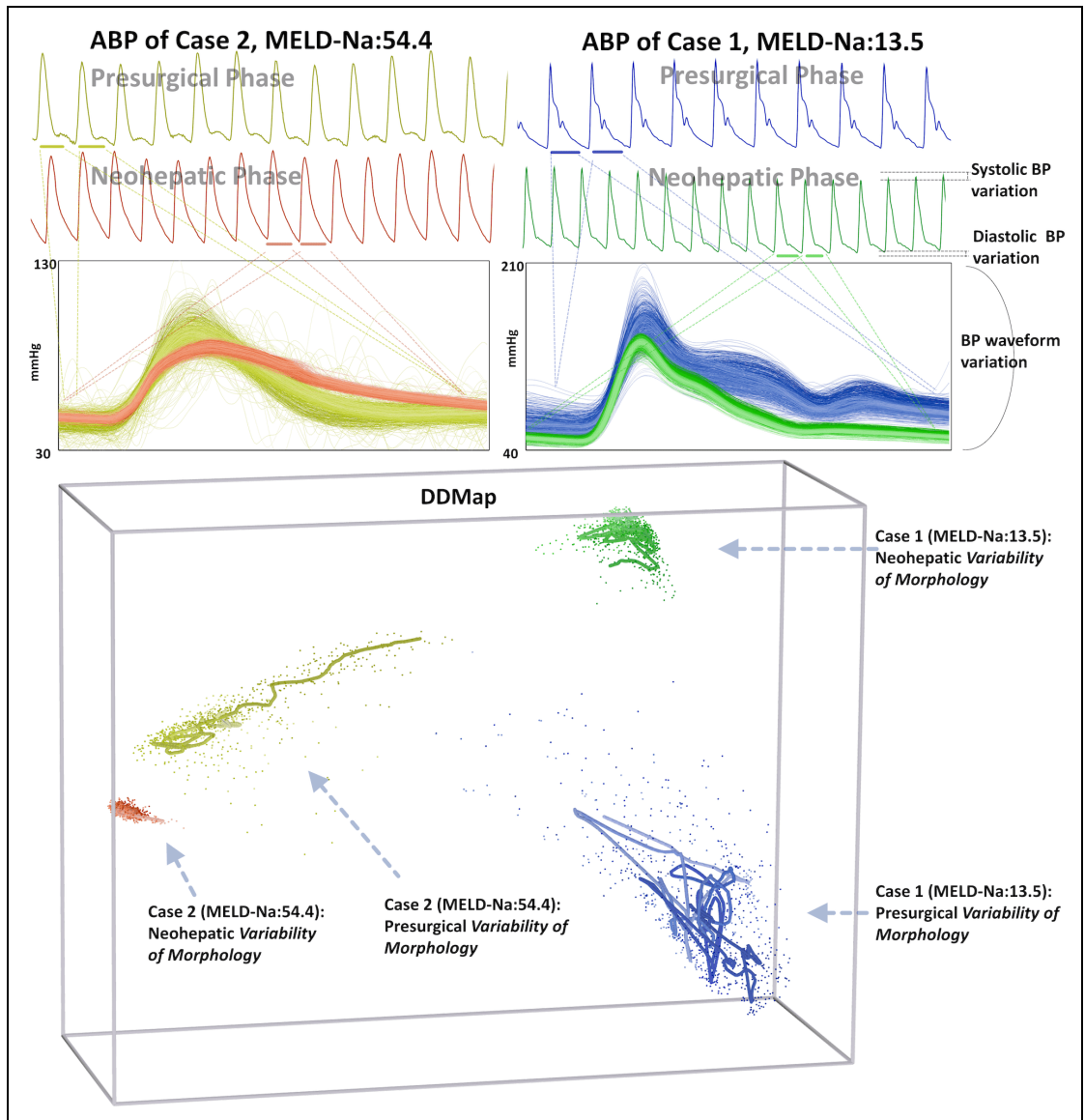

Fig. 1. DDMap 3D visualizations of the arterial blood pressure (ABP) waveforms during presurgical and neohepatic phases from 2 subjects for comparison. The variability of morphology from the original ABP beat-to-beat waveform data (upper panel and middle panel) is so subtle that it requires DDMap to grasp the inner structure. The evolution of the pulse cycle is shown in the form of 3D point cloud and their moving average trend (thick lines), which represent complex pathophysiological interactions. The fading in color indicates the direction of time while the lines represent the trends of the beat-to-beat fluctuation. In the presurgical phase, the variability of morphology is larger in case 1 (MELD score: 13.5) compared with that in case 2 (MELD score: 54.4). In both cases, the variabilities were decreased in the neohepatic phase. Note that the four trajectories and their trends present complex 3D shapes, which are simplified representations of the underlying variability of morphology in high dimensional space (multivariate).

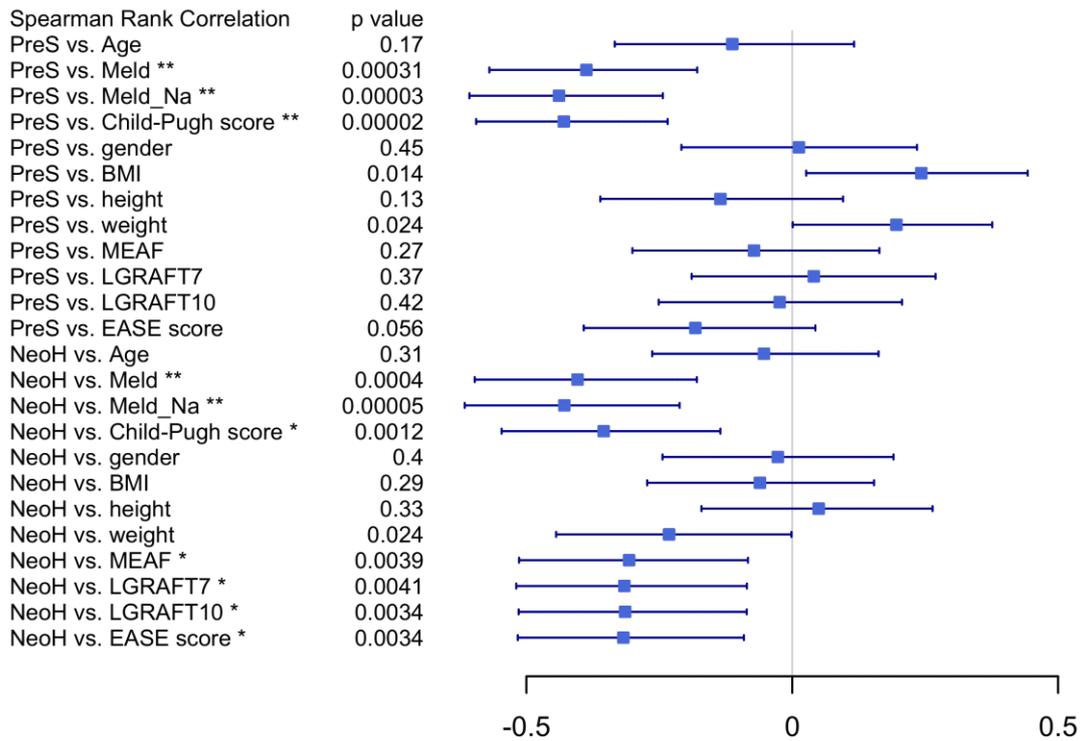

Fig. 2. Spearman's rank correlation between clinical data and the variability of morphology in the presurgical (PreS) and neohepatic (NeoH) phases, presented in terms of correlation coefficients, 95% confidence intervals (error bars), and p values (* <0.005, ** <0.0005). In the presurgical phase, the correlations with the three liver disease acuity scores were statistically significant. In the neohepatic phase, the complexity index was correlated with all four EAF scores (MEAF score, L-GrAFT$_7$, L-GrAFT$_{10}$, and EASE) as well as the MELD-Na and other scores. Note that a lower MELD score indicates better clinical conditions, such that the ideal correlation coefficient is −1.

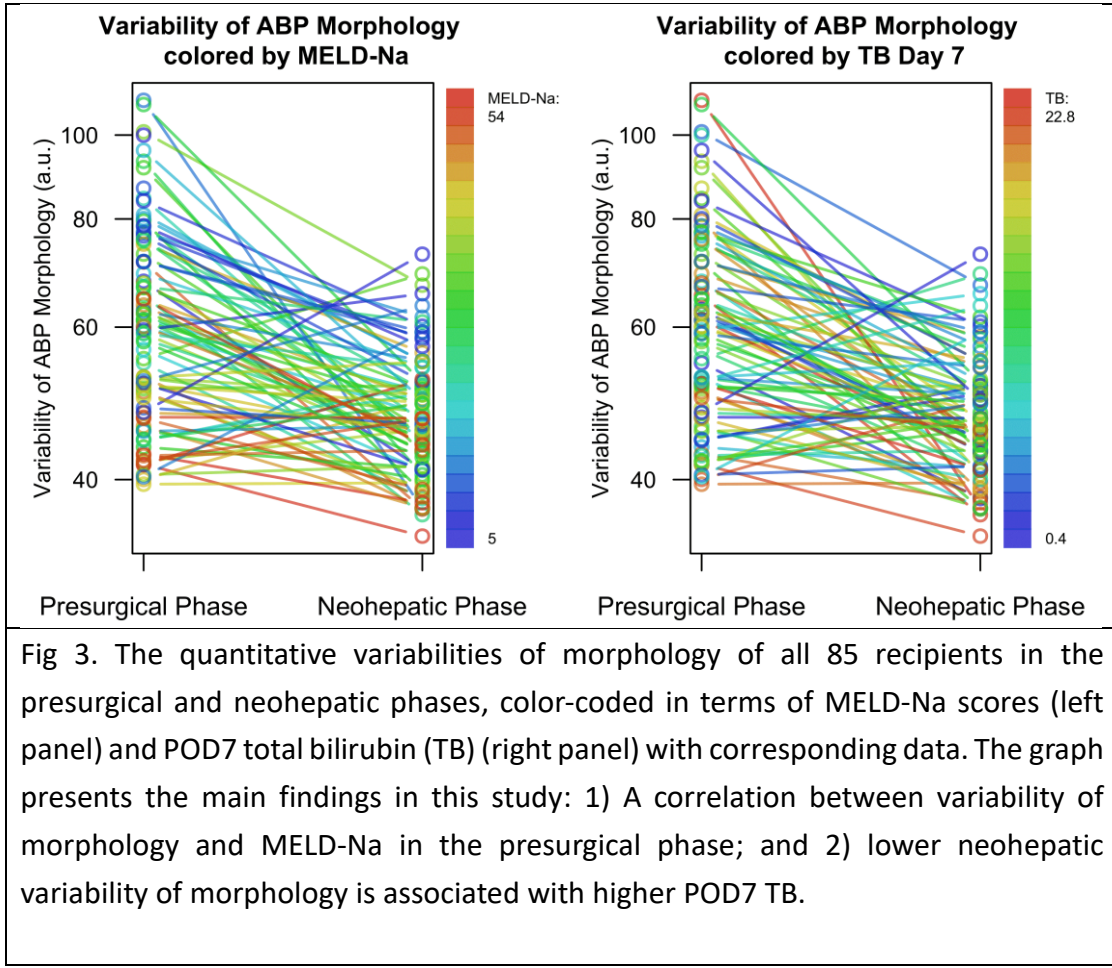

Fig 3. The quantitative variabilities of morphology of all 85 recipients in the presurgical and neohepatic phases, color-coded in terms of MELD-Na scores (left panel) and POD7 total bilirubin (TB) (right panel) with corresponding data. The graph presents the main findings in this study: 1) A correlation between variability of morphology and MELD-Na in the presurgical phase; and 2) lower neohepatic variability of morphology is associated with higher POD7 TB.

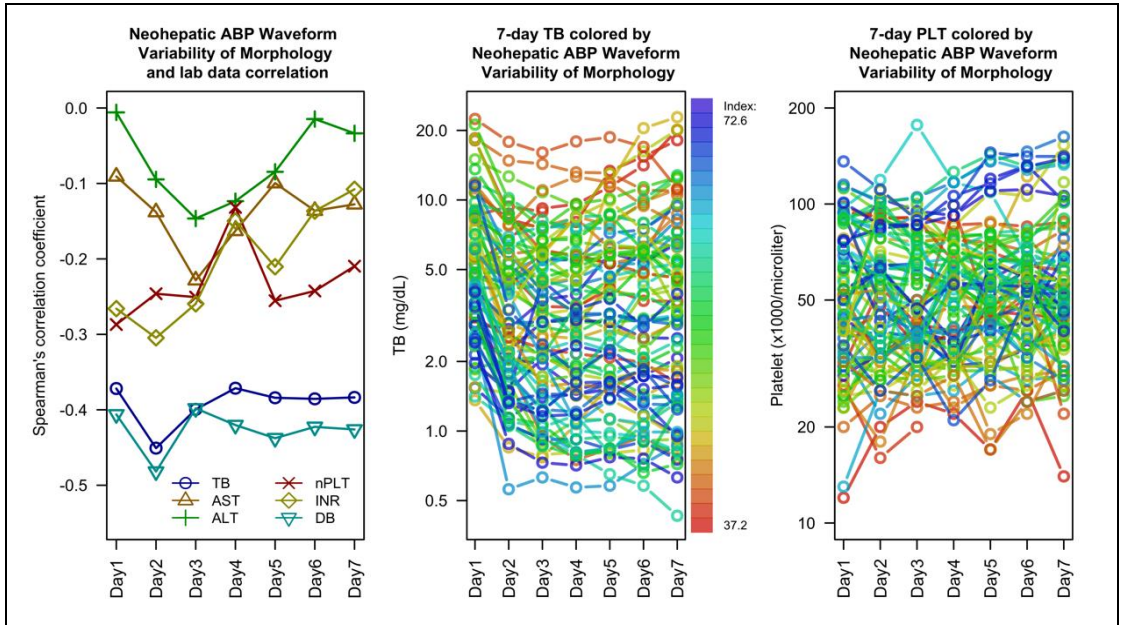

Fig.4. Correlation between the variability of morphology in the neohepatic phase and post-operative laboratory data using Spearman's correlations (left panel) or individual day1-to-day7 tracings of total bilirubin (middle panel) or platelet count (right panel). Note that the values are color-coded according to the rank of the neohepatic variability of morphology. Total bilirubin (TB), direct bilirubin (DB), and platelet count (nPLT representing negative platelet for a better visualization) provided the strongest correlations. Individuals of higher neohepatic variability of morphology (blue lines) showed a lower TB and a higher platelet count. ALT: alanine aminotransferase; AST: aspartate aminotransferase; INR: international normalized ratio.

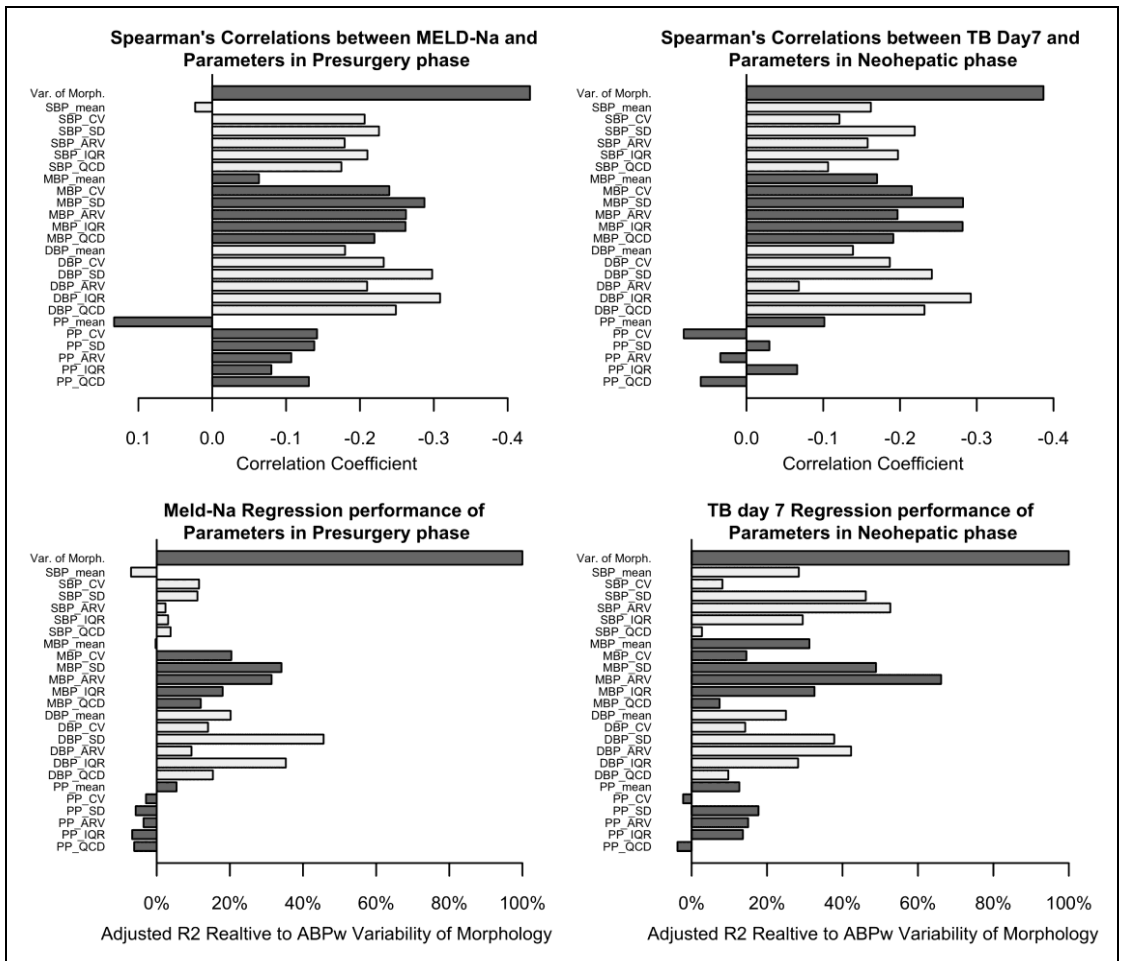

Fig.5. Performance comparison between variability of morphology of ABP waveform and the parameters derived from intraoperative available vital signs via the Spearman's correlation and the $R^2$ value of the regression analysis. With regard to MELD-Na score and total bilirubin (TB) postoperative day 7, variability of morphology performs best among the mean value and other blood pressure variation measures (CV: coefficient of variation; SD: standard deviation; ARV: average real variability; IQR: interquartile range; QCD: Quartile coefficient of dispersion) of the common blood pressure measures (SBP: systolic BP; MBP: mean BP; DBP: diastolic BP, PP: pulse blood pressure) plotted in different gray scale color as a visual aid. Note that an adjusted $R^2$ value in univariate linear regression is allowed to be negative.

Table 1. The demographic data. HBV, hepatitis B virus; HCV, hepatitis C virus; HCC,

| Characteristics | Number (%) | Median (IQR) | Mean (SD) |
|---|---|---|---|
| Recipient data | | | |
|   Age, y | NA | 59 (52–62) | 57 (8.5) |
|   Male sex | 60 (70.5) | NA | NA |
|   Main indication | | | |
|     HBV | 49 (58.3) | NA | NA |
|     HCV | 8 (9.5) | NA | NA |
|     Alcoholic cirrhosis | 12 (14.3) | NA | NA |
|     Other | 10 (11.9) | NA | NA |
|     HCC | 46 (54.8) | NA | NA |
|   MELD | NA | 18 (11–29) | 20.4 (11.5) |
|   MELD-Na | NA | 20 (14–32) | 22.4 (11.1) |
|   GRWR, % | NA | 1.07 (0.88–1.27) | 1.13 (0.35) |
|   CIT, min | NA | 58.5 (41–91.75) | 71.3 (40.8) |
|   WIT, min | NA | 40 (32–52.25) | 42.7 (14.2) |
| Graft data | | | |
|   Type | | | |
|     RL without MHV, con(-) | 43 (50.6) | NA | NA |
|     RL without MHV, con(+) | 29 (34.1) | NA | NA |
|     RL with MHV | 3 (3.5) | NA | NA |
|     LL with MHV | 10 (11.8) | NA | NA |
|   Fatty liver, macrovesicular, % | | NA | NA |
|     None | 39 (45.9) | NA | NA |
|     1~10 | 36 (42.4) | NA | NA |
|     11~20 | 7 (8.2) | NA | NA |
|     21~30 | 3 (3.5) | NA | NA |

hepatocellular carcinoma; MELD, Model for End-stage Liver Disease; GRWR, graft to recipient weight ratio; CIT, cold ischemic time; WIT, warm ischemic time; RL, right liver lobe; MHV, middle hepatic vein; LL, left liver lobe; con(-)/(+), without/with MHV territory reconstruction; NA, not applicable.

# Supplementary materials

In the following, we describe the technical aspects of the *variability of morphology* of arterial blood pressure (ABP) waveform.

The subtle evolution of the ABP waveforms reflects various pathophysiological effects on the cardiovascular system. Dynamic diffusion maps (DDMap) allows us to capture the beat-to-beat morphological change in the high dimensional space by converting the sequential beats to a trajectory in 3-D (Fig. S1)[9]. The variability of morphology is quantified by the length of the moving average trend of the trajectory in the high dimensional space. Here the rational of the moving average is removing the fluctuation while revealing the underlying dynamics.

As the moving average trend represents the ABP waveform morphological dynamics on a wave-shape manifold, which is a high-dimensional structure that parametrizes the ABP morphology. The length of the trend grants the quantitative variability of morphology, which is calculated as $Variability\ of\ Morphology = Trend\ length/pulse\ count$ . After obtaining the quantitative index, we then standardize it with respect to other reference trend data.

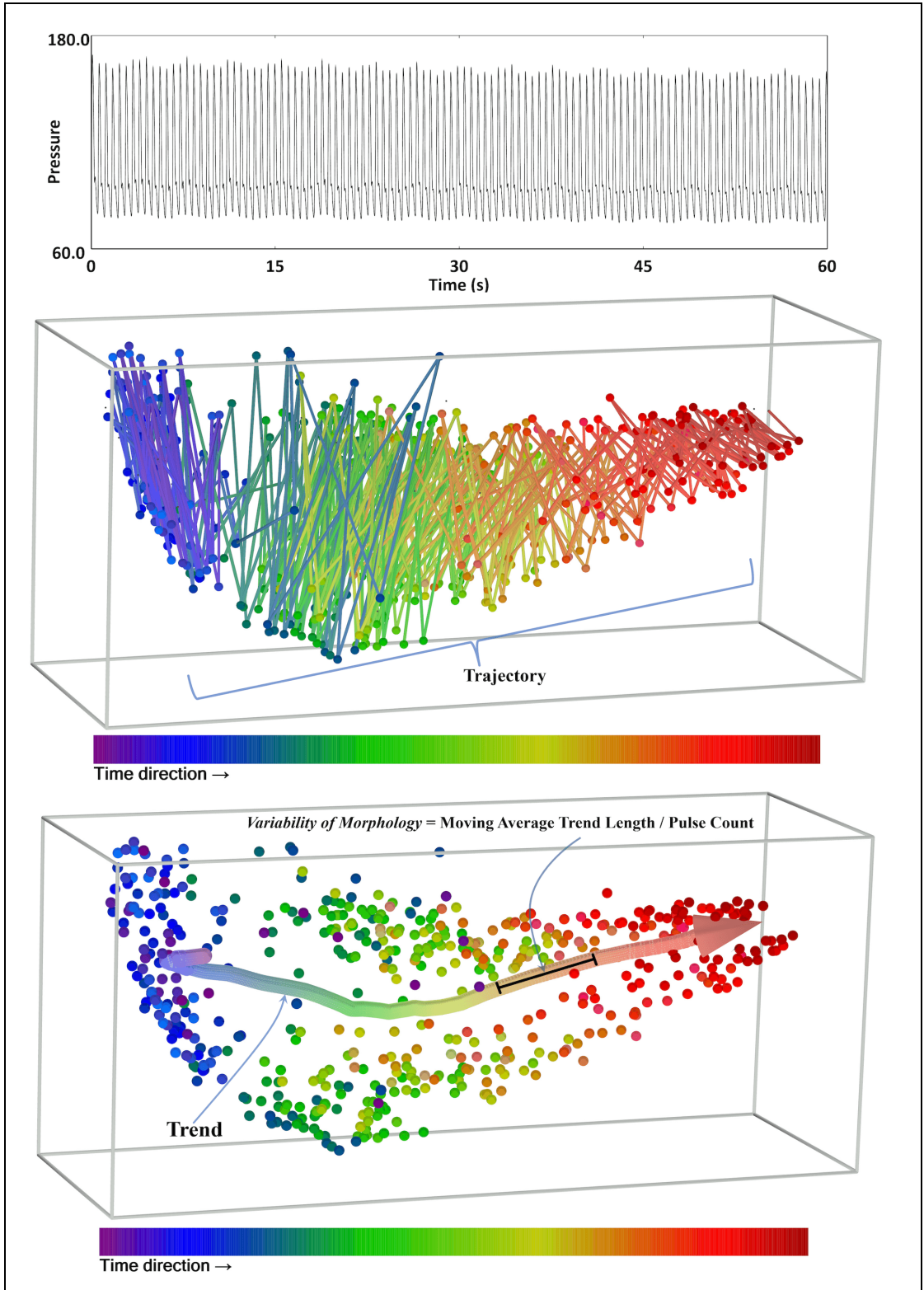

Fig. S1. The evolution of a waveform is difficult to extract via bare eye inspection from a 1-minute tracing (upper), whereas 5-minute data reveal a pulse-by-pulse trajectory (middle). Note that this 3-D image was created via DDMap. The *variability of morphology* provides additional information pertaining to the shape of the trajectory in the form of quantitative moving average trend (lower) in this high-

> dimensional space. The middle panel and lower panel present the same data in different format to help establish the relation between the trajectory and the variability of morphology.

## DDMap Algorithm

### Step 1: Processing ABP Waveforms

Pulsation driven by the heartbeat is a universal feature of nearly all cardiovascular signals. Recorded ABP waveforms can be denoted as $x^A \in \mathbb{R}^N$, where $N$ indicates the number of sampling points. In the current study, the sampling rate was 300 Hz; that is, the signal lasts for $N/300$ seconds since the sampling rate is 300Hz. The first step in generating a data point from every pulse involves segmenting $x^A$ into pulses. The maximum of the first derivation of the ascending part of every ABP waveform is used as its fiducial point. We denote the $i$-th fiducial point as $n_i$ and assume that there are $n$ intact pulse cycles in $x^A$. Segmentation results in a collection of consecutive pulses, where the $i$-th segment represents one pulse cycle as $\bar{x}_i^A := [x^A(n_i), x^A(n_i+1), \ldots, x^A(n_i+q-1)]^T$, where $q$ is the minimal length of all segments to which each pulse is truncated to ensure uniform size. Denote $\chi^A$ to be the set of $n$ pulse cycles, $\bar{x}_1^A, \ldots, \bar{x}_n^A$. Technically, it is assumed that $\chi^A$ can be parameterized by a low dimensional manifold. The subsequent task involves using the manifold learning algorithm to unveil the hidden dynamic structure.

### Step 2: Dynamic Diffusion Map (DDMap) and Variability of Morphology

The *Dynamic Diffusion map* (DDMap) was used to study the complex structure of $\chi^A$. As a variation of the commonly applied unsupervised learning algorithm diffusion maps (DM)[34], DDMap was designed for the analysis of ultra-long-term biomedical signals. In the current study, we chose DDMap because it can preserve the global structure with rigorous theoretical support. We summarize the algorithm below and refer readers to the literature for technical details[10, 11].

The first step involves constructing an $n \times n$ affinity matrix $W$ from $\chi^A$ comprising of $n$ pulse cycles, such that $W_{ij} = e^{-|\bar{x}_i^A - \bar{x}_j^A|^2/\varepsilon}$, where $i, j = 1, \ldots, n$ and bandwidth $\epsilon > 0$ is set as the 25% percentile of all pairwise distances. From the learning perspective, $W_{ij}$ quantifies the dissimilarities of the pair $x_i$ and $x_j$; that is, the value of $W_{ij}$ is inversely proportional to the degree of similarity between the two points. Next, we construct an $n \times n$ degree matrix $D$, such that $D(i,i) = \sum_{j=1}^{n} W(i,j)$, where $i = 1, \ldots, n$. Using matrices $W$ and $D$, we obtain the transition matrix $A := D^{-1}W$, which depicts a random walk in the point cloud $\chi^A$. Denote the eigen-decomposition of $A$ as $A = U\Lambda V^T$, where $\Lambda$ contains eigenvalues $1 = \lambda_1 > \lambda_2 \geq \lambda_3 \geq \cdots \lambda_n$ of $A$ and $U$ contains the corresponding right eigenvectors $\phi_1, \phi_2, \ldots, \phi_n \in \mathbb{R}^n$. Finally, we obtain the DDMap as $\Phi_t: \bar{x}_j^A \mapsto \left(\lambda_2^t \phi_2(j), \lambda_3^t \phi_3(j), \ldots, \lambda_{\hat{d}+1}^t \phi_{\hat{d}+1}(j)\right) \in \mathbb{R}^{\hat{d}}$, where $j = 1, \ldots, n$, $t > 0$ is the user-defined *diffusion time*, and $\hat{d}$ is the user-defined embedding dimension. Note that $j$ encodes the temporal relationship among pulse cycles, and the sequence of points $\Phi_t(\bar{x}_1^A), \ldots, \Phi_t(\bar{x}_n^A)$ could be linked following this temporal relationship and form a trajectory in the $\hat{d}$-dim space. This trajectory describes how the pulse cycle morphology changes from one to another. We discard $\lambda_1$ and $\phi_1$, as they contain no information, and set diffusion time $t$ to 1 in order to obtain the "standard" scale of the image. If $\hat{d}$=3, then DDMap converts the pulse cycles into a 3-D scatterplot image, which allows users to visualize an overall picture of the dataset. To quantify the nonlinear dynamic structure encoded in the evolving pulse cycle, we empirically set dimension $\hat{d}$ as 15, due to the fact that the eigenvalues decay rapidly enough that the eigenvalues of $\hat{d} > 15$ can simply be neglected. In other words, DDMap converts the pulse cycle into a vector of length 15.

In the quantification of *variability of morphology*, we first define the *diffusion distance* (DDist) between two points $\bar{x}_i^A$ and $\bar{x}_j^A$, which is the straight line distance defined as $D_t(\bar{x}_i^A, \bar{x}_j^A) := \left\|\Phi_t(\bar{x}_i^A) - \Phi_t(\bar{x}_j^A)\right\|_{l^2}$. Note that DDist is used to quantify the degree of similarity between data points. Until now it is worth mentioning that from the DDist we can see the difference between Diffusion map algorithm and the famous Principal Component Analysis (PCA) algorithm despite both use the eigendecomposition. PCA cannot provide similar property of the DDist. The *variability of morphology* is defined as follows using DDist. First, we apply a moving median followed by a moving mean filter to extract the trend, $\widehat{T}_j := \frac{1}{k}\sum_{m=j-k+1}^{j} Median\left(\Phi_t(\bar{x}_m^A), \Phi_t(\bar{x}_{m-1}^A), \ldots, \Phi_t(\bar{x}_{m-k+1}^A)\right)$, where $j = k, \ldots, n$ and the filter is parameterized by the window length $k$. The aim of *variability of morphology* is to quantify the rate with which the trend evolves, which is defined by the formula $variability\ of\ morphology = (\sum_{i=k+1}^{n}\|\widehat{T}_i - \widehat{T}_{i-1}\|_{l^2})/T$, where $T$ is the length of the ABP signal (in units of seconds) and $n$ is the number of pulse cycles. For the convenience in future clinical applications, it would be preferable for the *variability of morphology* to be around the 0-100 range. To this end, we pool all ABP data segments (presurgical phase: 85 cases, neohepatic phase: 83 cases) and use the following formula to obtain the final *variability of morphology*:

*Final variability of morphology of the ith subject*

$$= \frac{\textit{Variability of Morphology of the ith subject} - \textit{median}(\textit{pooled Variabilities of Morphology})}{\textit{IQR}(\textit{pooled Variabilities of Morphology})} \times 25 + 60.$$

### Sensitivity analysis of MELD score correlations

To elucidate the robustness of the algorithm, we investigated the moving average window length *k* in the following sensitivity analysis. Sensitivity analysis here involves testing the effect of filter size parameter K, which determines the moving average behavior of the trend. The objective is to determine whether the *variabilities of morphology* are sensitive to the selected parameters. In an analysis based on the Spearman correlation between presurgical ABP data and MELD scores (Fig. S2), the *variability of morphology* achieved consistent correlation from a wide range of window size parameters.

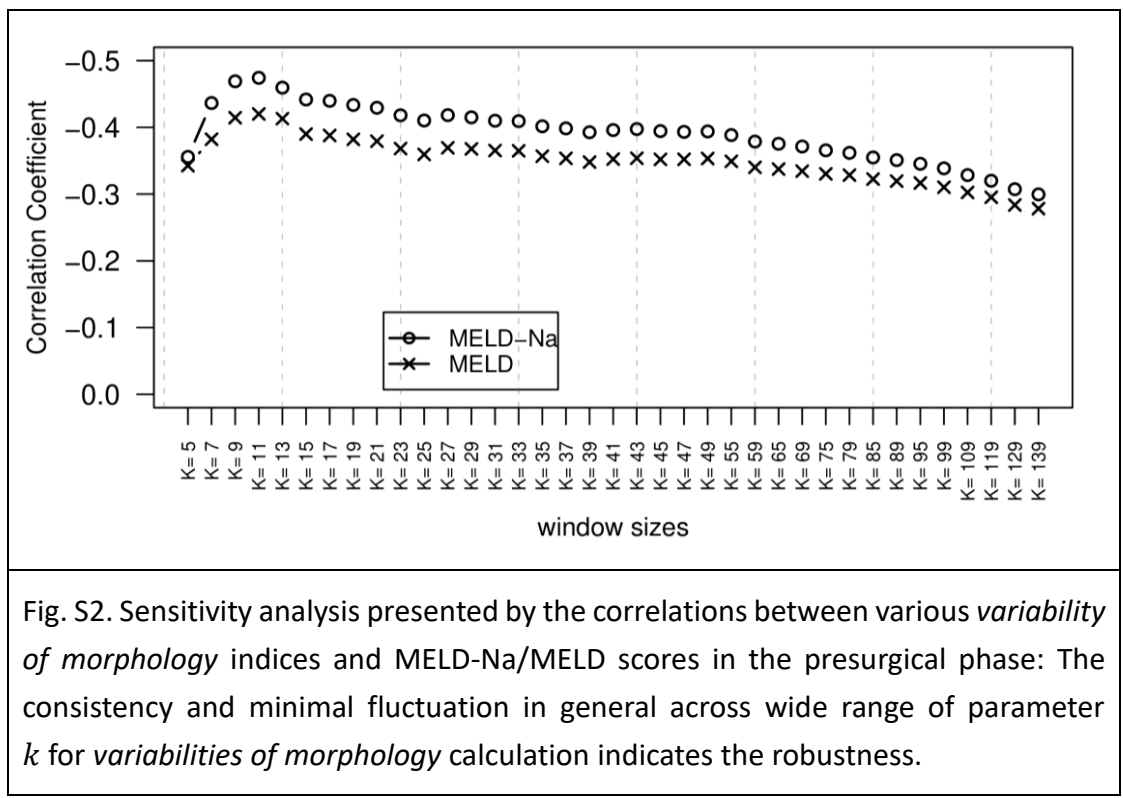

Fig. S2. Sensitivity analysis presented by the correlations between various *variability of morphology* indices and MELD-Na/MELD scores in the presurgical phase: The consistency and minimal fluctuation in general across wide range of parameter $k$ for *variabilities of morphology* calculation indicates the robustness.

### Mean blood pressure as the predictor

In this sensitivity analysis (Fig. S3), we substitute the variability of morphology by the mean blood pressure (MBP) and evaluate the degree of correlation between MBP and MELD scores. In the presurgical phase, compared with the correlations between the *variability of morphology* and MELD-Na and MELD, the correlations between MBP and MELD-Na and MELD are lower (MELD-Na: -0.439 → -0.174, MELD: -0.387 → -0.150).

Sensitivity Test

One validation test was a surrogate data test involving the random permutation of the time sequence of ABP pulses. We observed decreases in the correlations pertaining the *variability of morphology* index (MELD-Na: -0.439 → -0.287, MELD: -0.387 → -0.266) with statistical significance. It is worth mentioning that the time sequence information contributes the performance of the *variability of morphology*, while the usual variance measures in descriptive statistics, such as the standard deviation, coefficient of variation, or the interquartile range, do not take time sequence information into account.

Another validation involved the permutation of case labeling of random pair MELD scores and ABP waveforms. This loss of association supports our hypothesis pertaining to a link between the *variability of morphology* and disease acuity.

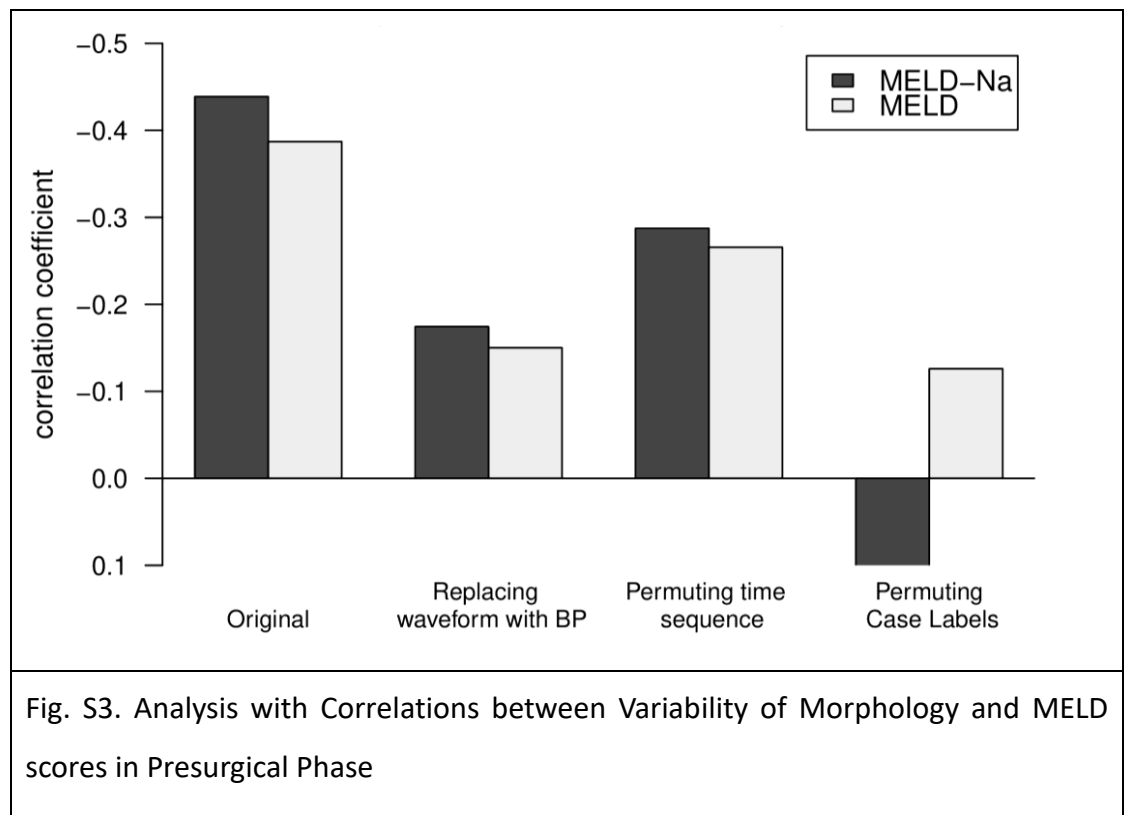

Fig. S3. Analysis with Correlations between Variability of Morphology and MELD scores in Presurgical Phase

The final sensitivity analysis involves the reduction of the duration of the variability of morphology calculation (Fig. S4). The result shows slightly impact of the correlation in comparison of 10 minutes duration, which also supports the robustness of the variability of morphology.

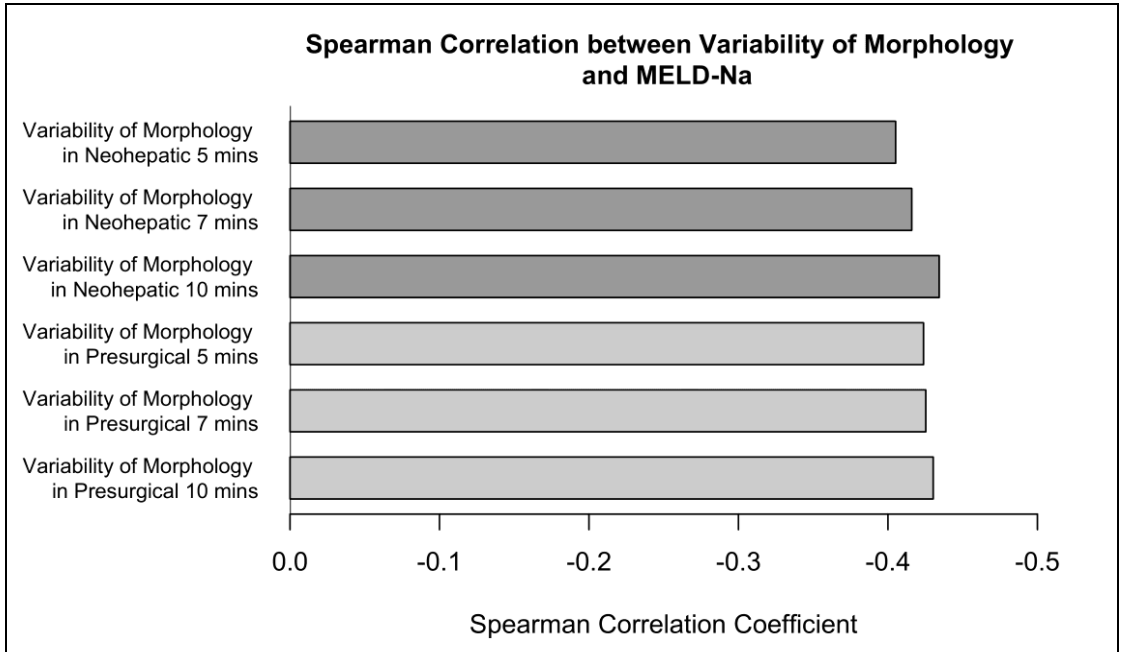

Fig. S4. Reduced time duration (7 minutes and 5 minutes) shows limited effect on the performance variability of morphology in comparison with 10 minutes duration